\documentstyle[epsfig]{JPP}
\begin{document}
%Switch figures on and off
%\newcommand{\FIG}[1]{}
\newcommand{\FIG}[1]{#1}
\newcommand{\PREP}[1]{#1}

\newcommand{\vv}{\mbox{\bf v}}
\newcommand{\BB}{\mbox{\bf B}}
%
% Instellen float fractions
%
\renewcommand{\floatpagefraction}{.9}
\renewcommand{\textfraction}{.1}
\renewcommand{\topfraction}{.9}
\renewcommand{\bottomfraction}{.9}
\title[Kelvin-Helmholtz instability with magnetic fields]
{Growth and saturation 
of the Kelvin-Helmholtz instability with parallel and anti-parallel
magnetic fields}

\author[R. Keppens, G. T\'oth, R.H.J.~Westermann, J.P.~Goedbloed]
{R\ls O\ls N\ls Y\ns K\ls E\ls P\ls P\ls E\ls N\ls S$^1$\ns, 
G. T\ls \'O\ls T\ls H$^2$\ns, 
R.H.J.~W\ls E\ls S\ls T\ls E\ls R\ls M\ls A\ls N\ls N$^3$\ns, 
\and\ns  J.P.~G\ls O\ls E\ls D\ls B\ls L\ls O\ls E\ls D$^1$\ns}

\affiliation{$^1$FOM-Institute for Plasma-Physics Rijnhuizen, \\
P.O. Box 1207, 3430 BE Nieuwegein, The Netherlands, \\
keppens@rijnh.nl goedbloed@rijnh.nl \\[\affilskip]
$^2$Department of Atomic Physics, E\"{o}tv\"{o}s University, \\
Puskin u. 5-7, Budapest 1088, Hungary, \\
gtoth@hercules.elte.hu \\[\affilskip]
$^3$Astronomical Institute, Utrecht University, \\
P.O. Box 80000, 3508 TA Utrecht, The Netherlands, \\
westermn@fys.ruu.nl 
}

\date{?? February 1998}
\maketitle

\begin{abstract}

We investigate the Kelvin-Helmholtz  
instability occuring at the interface of
a shear flow configuration in 2D compressible magnetohydrodynamics (MHD). 
The linear growth and the subsequent non-linear saturation of the instability 
are studied numerically.
We consider an initial magnetic field aligned with
the shear flow, and analyze the differences between cases where the initial
field is
unidirectional everywhere (uniform case), and where the field changes sign
at the interface (reversed case).
We recover and extend known results for pure hydrodynamic and MHD cases 
with a discussion of the dependence of the non-linear saturation on
the wavenumber, the sound Mach
number, and the Alfv\'enic Mach number for the MHD case. 

A reversed field acts to destabilize the linear phase of the
Kelvin-Helmholtz instability compared to the pure hydrodynamic case, 
while a uniform field suppresses its growth. 
In resistive MHD, reconnection events almost instantly accelerate the buildup
of a global plasma circulation. They play an important role throughout the 
further non-linear evolution as well, since the initial current sheet
gets amplified by the vortex flow and can become unstable
to tearing instabilities forming magnetic islands. As a result, the saturation
behaviour and the overall evolution of the density and the magnetic field
is markedly different for the uniform versus the reversed field case.

\end{abstract}

\section{Introduction}

The Kelvin-Helmholtz (KH) instability occurs at the interface between two 
fluids or plasmas moving in opposite directions. Pioneering studies
were made by Chan\-dra\-se\-khar (1961).
%~\cite{chandra}. 
As shear flows are
present in many astrophysical situations, the KH instability receives
continuous attention in the astrophysics literature. In solar coronal loops,
shear flows arise in the vicinity of a resonant radius when the magnetized
loop is perturbed at a frequency which matches
the characteristic Alfv\'{e}n frequency at this radius. Recent
studies address whether these shear flows are KH unstable 
(\cite{karpen};~\cite{poedts}).
In the heliosphere, the solar wind flows past planetary magnetospheres,
and instabilities may occur at bow shocks or further out in the 
magnetotails (see, e.g.~\cite{uberoi}). Similarly, KH instabilities
are studied at the heliopause, where the solar wind is halted and 
meets the interstellar medium (\cite{chalov}).
Numerous observations of extragalactic jets inspired 
research into the KH instability taking relativistic effects into account.
A recent example is found in~\cite{hanasz}.

We investigate both the linear and the non-linear regime of the KH
instability for compressible hydrodynamic (HD)
and magnetohydrodynamic (MHD) cases. Our numerical
study makes use of the Versatile Advection Code (VAC) 
for solving the fluid equations, developed by T\'oth (1996, 1997).
This
code can be used as a convenient tool to handle hydrodynamic and 
magnetohydrodynamic one-, two-, or three-dimensional problems in 
astrophysics (see~\cite{vacporto}). 
We perform all our calculations in two dimensions. 

First, we briefly summarize important
findings preceding and augmenting our study.
Theoretical studies of the KH instability started with the linear stability
analysis of~\cite{chandra} for incompressible HD and
MHD cases. It was noted how a uniform magnetic field, parallel to the
shear flow, completely stabilizes the KH instability when the velocity
jump across the shear layer is less than twice the Alfv\'{e}n speed. 
\cite{blumen} extended the 
linear study to the hydrodynamic compressible case and found instability
when the sound speed exceeds half the velocity jump. In our investigation,
we will therefore restrict attention to cases where half the total velocity
jump is below the sound speed, but above the Alfv\'{e}n speed.
The extension of these linear studies to compressible MHD cases is
found in~\cite{miupr}. While the magnetic field 
was taken to be uniform, they allowed for magnetic fields and wavenumbers
with arbitrary orientations in the plane perpendicular to the velocity gradient.
Since we restrict ourselves to 2D configurations, our linear results
for uniform magnetic fields parallel to the shear flow recover
their findings. In particular, we similarly study variations with
wavenumber, sound Mach number and magnetic field strength. We extend their
findings by discussing the further non-linear evolution as well.  
In addition, we simulate the KH instability when the initial 
field is anti-parallel (or reversed) in both opposite flowing plasma layers. 

Recently,~\cite{frank} carried out non-linear
2D MHD calculations for two cases. They took the field parallel to 
the shear flow and compared the KH evolution of a `weak' field case
with a `strong' field case. The velocity jump across the shear layer 
equaled the sound speed, and amounted to 5 and 2.5 times 
the Alfv\'{e}n speed, respectively. It was found that the magnetic tension
also stabilizes the non-linear regime for these two
cases. We confirm and extend this result for the uniform field
over a wider range of Alfv\'{e}n and sound Mach numbers. 
\cite{frank} found that 
numerical dissipation, mimicking viscous and resistive effects,
eventually led to similar end states consisting of a stable
laminar flow with dynamically aligned velocity and magnetic fields.

A systematic numerical study of the 2D uniform case in MHD was done
by Ma\-la\-go\-li et al. (1996).
%%~\cite{malag}. 
They varied the ratio of the
Alfv\'en speed to the sound speed, and could generally
identify three stages in the instability. They referred
to these stages as the linear stage, the dissipative
transient stage with intermittent reconnection events, and the saturation
stage where small scale turbulent motions decay to form aligned structures.
Their final stage is qualitatively similar to the laminar flow end state of
\cite{frank}. We focus attention to the two stages {\em
preceding} turbulence, where the amplification and the dynamic
influence of the magnetic field grows and eventually halts the KH instability.

The main extension presented here is a consideration
of the reversed field case in 2D compressible MHD. There,
new physical effects emerge as reconnection can occur earlier in the 
evolution of the KH instability. We compare this reversed case 
with the pure hydrodynamic and uniform magnetohydrodynamic case.
Independent investigations by~\cite{dahl97} studied similar
`current-vortex' sheets where both the initial velocity and the magnetic
field are given by hyperbolic tangent profiles. 
By considering relatively strong fields, these authors
could investigate resistive instabilities, modified by the KH flow.
We show how the KH instability in the presence of an initially weak reversed
field can induce tearing instabilities. Their study was done
in incompressible visco-resistive MHD, so we extend their results by
adding compressibility. 
Another important difference is that we take a discontinuous field
reversal at the shear flow interface, while \cite{dahl97} took
the ratio of the magnetic shear width to the velocity shear width 
around unity. We have therefore investigated the dependence on the 
initial magnetic field configuration. This also allows a more detailed
comparison with the incompressible studies of `current-vortex' sheets.
The most recent study by
\cite{dahl98} considers nonlinear effects analytically,
under the assumptions of incompressibility, while neglecting higher harmonics
and the distortion of the fundamental disturbance.
Our numerical study allows
us to relax these assumptions to fully 2D compressible situations.

The manuscript is built up as follows:
in section~\ref{s-B} we list the conservation laws, summarize the initial 
conditions, and we give a brief discription of the KH instability. 
The numerical method is outlined in section~\ref{s-N}.
The results are split into a discussion of the linear regime 
(section~\ref{s-LR}) and an in depth study of the non-linear
saturation behaviour (section~\ref{s-NLR}).
We end with a discussion and conclusions in section~\ref{s-C}.

\section{Conservation laws and initial configuration}\label{s-B}

The MHD equations are written in conservation form.  
The conservative variables are density $\rho$, momentum $\rho \vv$,
energy density $\cal{E}$, and the magnetic field $\BB$. 
The conservation of mass is simply written as
\begin{equation}
\frac{\partial \rho}{\partial t}+\nabla \cdot (\rho \vv)=0.
\end{equation}
The evolution equation for the momentum density $\rho \vv$ reads
\begin{equation} 
\frac{\partial (\rho \vv)}{\partial t}+ \nabla \cdot [ \rho \vv \vv + p_{tot} I-\BB \BB]=0,
\end{equation}
 where $\rho \vv \vv$ is the Reynolds stress tensor, 
$p_{tot}=p + \frac{1}{2}B^2$ is the total pressure, 
and $I$ is the identity tensor. 
The thermal pressure $p$ is related to the energy density as 
$p=(\gamma-1)({\cal{E}}-\frac{1}{2}\rho v^{2}-\frac{1}{2}B^{2})$. 
We set the adiabatic gas constant $\gamma$ equal to $5/3$. 
The magnetic part of Maxwell's stress tensor is $\frac{1}{2}B^2I-\BB \BB$.
Magnetic units are defined such that the magnetic permeability is unity.
The induction equation is
\begin{equation}
\frac{\partial \BB}{\partial t}+ 
\nabla \cdot (\vv \BB-\BB \vv)= \eta \nabla^2 \BB.
\label{q-b}
\end{equation}
Ideal MHD corresponds to a zero resistivity $\eta$ and ensures that magnetic
flux is conserved. In resistive MHD, field lines can reconnect.
The energy equation we use is
\begin{equation}
\frac{\partial {\cal{E}}}{\partial t}+ \nabla \cdot ({\cal{E}} \vv) + \nabla \cdot (p_{tot} \vv) - \nabla \cdot (\vv \cdot \BB \BB )=
\nabla \cdot \left[\BB \times \eta (\nabla \times \BB)\right].
\label{q-e}
\end{equation}
In resistive MHD ($\eta >0$), the right hand side contains the
Ohmic heat term $\eta (\nabla \times \BB)^2$, which 
is a source of internal energy. 

We solve the above set of non-linear ideal MHD equations as an
initial value problem in two spatial dimensions.
%%, assuming plane parallel symmetry in the third dimension.
We consider a Cartesian two-dimensional rectangular grid with $0 \le x \le L_x$ 
and $-L_y \le y \le L_y$. We take $L_y$ as our unit of length. The initial 
pressure $p_0$ and density $\rho_0$ are set to unity throughout the domain
and thus define our normalization.

We distinguish between the following three cases: (i)
a purely hydrodynamic case with $B=0$ at all times; (ii) a `uniform' MHD
case where the magnetic field at $t=0$ is set to $B_0\hat x$ everywhere;
and (iii) a `reversed' MHD case where the magnetic field at $t=0$ is 
$-B_0\hat x$ for $y>0$ and $B_0\hat x$ for $y<0$.
The reversed case must be seen as a limiting case of a
continuous field reversal, where $B_x=B_0\tanh(y/b)$ and $b\rightarrow 0$.
We will explicitly discuss this connection with an initially smooth current
sheet. Note that when $b \neq 0$, we have a non-constant 
initial pressure $p(y)$ from total pressure equilibrium.

We apply a shear velocity in the $x$-direction with amplitude $V_0$, of
the form $v_x=V_0\tanh(y/a)$.
The width of the `inner' region where the velocity reverses is always set to
$a=0.05$. In order to guarantee instability (\cite{chandra}, 
\cite{blumen},~\cite{miupr})
we vary the amplitude of the shear such that
\begin{equation}
B_0/\sqrt{\rho_0} \equiv v_a<V_0<c_s \equiv \sqrt{\gamma p_0/\rho_0}.
\end{equation}
The sound speed $c_s$ and the Alfv\'{e}n speed $v_a$ 
are used to define the sound Mach number $M_s\equiv V_0/c_s$ and
the Alfv\'{e}nic Mach number $M_a\equiv V_0/v_a$, respectively.
We perturb this initial configuration with a velocity component perpendicular 
to the background shear velocity of the form 
$\delta v_y=v_{y0}\sin(k_x x)\exp{[-y^{2}/\sigma^{2}]}$.
We take the amplitude of this velocity perturbation $v_{y0}=0.0001$, 
which is always much smaller than the shear velocity $V_0$. 
The Gaussian component of the velocity perturbation $\delta v_y$ has a 
parameter $\sigma$, representing the decay of the amplitude $v_{y0}$ in 
the outer region $|y|>a$, and the ratio $\sigma/a$ is set to 4 in our 
calculations. 

\begin{figure}[ht]
 \FIG{\vspace{9cm}
 \includegraphics{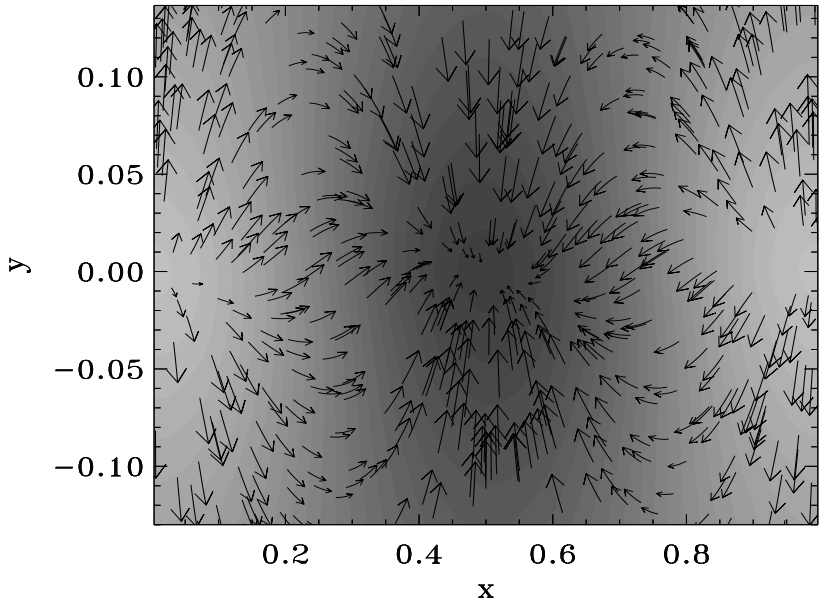}}
 \caption{The structure of the KH instability. The inner region containing the
background shear velocity is $|y|<a=0.05$.  
We show the density perturbation resulting from the initial velocity
perturbation (bright is higher density, dark corresponds to lower density).
Also shown is the induced pressure gradient force field which enhances the
initial vertical displacement in the centre of the layer.}
\label{f-khstruc}
\end{figure}

With these initial conditions and the use of periodic boundary conditions
in the $x$-direction, we can simulate the development of
a KH instability at certain Mach number and wavenumber $k_x=2\pi/L_x$.
The driving force behind the KH instability and its resulting
spatial structure is easily understood from Fig.~\ref{f-khstruc}
(see also the schematic Fig.~2 in~\cite{miupr}).  

Consider first what happens in the
inner region $|y| < a$ containing the background shear velocity.
When we impose a sinusoidal perturbation $\delta{v_y} \sim \sin (k_x x)$ 
at $y=0$, the velocity shear $v_x' = d v_x /dy$ in the equilibrium
flow produces a force $-\rho v_x'\delta v_y$ in the $x$-direction, such that
acceleration of the plasma in the peaks of the sine wave is
anti-parallel to the one felt in the throughs.
The resulting density perturbation is a periodic depletion 
and enhancement as shown in Fig.~\ref{f-khstruc},
which is horizontally displaced from the imposed
velocity perturbation. The pressure perturbation follows the
density perturbation, and sets up a pressure gradient force that
enhances the initial vertical displacement within $|y| < a$.
The pressure gradient force is indicated by the arrows in Fig.~\ref{f-khstruc}.
Note how in the inner region, the pressure gradient is in phase with the
initial velocity disturbance (from bottom to top in the left half of
the figure, and from top to bottom in the right half).
Hence, the situation is inherently unstable.

The situation is different in the outer region $|y| > a$.
There, the imposed sinusoidal perturbation is damped by 
$\exp{[-y^{2}/\sigma^{2}}]$, and the background flow is essentially uniform.
The pressure gradient force is out of phase with the initial flow
perturbation. The net result is a clock-wise
plasma circulation around the central density depletions. 
Matter pushed out by the excess pressure created in the central layer
which reaches $y=+a$ gets entrailed to the right by the background flow 
and is pulled towards the central layer again at the throughs of the initial
perturbation.

\section{Numerical method}\label{s-N}

All calculations are performed with the Versatile Advection Code 
(VAC, see T\'{o}th (1996, 1997)).
%%~\cite{vac},~\cite{vacvienna}). 
VAC is a general code for solving systems of conservation laws,
as e.g., the HD and MHD equations. Although several spatial and 
temporal discretizations are implemented in VAC, we consistently use
the explicit TVD-scheme, 
which is a one-step Total Variation Diminishing (TVD) 
scheme employing a Roe-type approximate Riemann solver,~\cite{roe}. 
This shock-capturing, second order accurate scheme limits the jumps
allowed in each of the characteristic wave fields to ensure the TVD property.
In the uniform MHD cases, we used the slightly diffusive {\it minmod} limiter.
Our method is then essentially identical with the one used by~\cite{frank}.
For all hydrodynamic and reversed MHD cases, we used the sharper 
{\it Woodward} limiter (see~\cite{toth-odstrcil}).
For the uniform case, the use of this limiter occasionally caused
numerical problems. However, the evolution of the uniform case in ideal MHD is not
much affected by the actual choice of limiting, as long as numerical
dissipation does not play a role.
We apply a projection scheme at every time step to remove any numerically
generated divergence of the magnetic field. 

Previous authors have considered very
high grid resolutions. \cite{frank} went up to
$512 \times 512$ cells,~\cite{malag} 
used $256 \times 512$ cells.
To investigate the linear growth phase and determine in which way the
equilibrium parameters influence the non-linear saturation, 
we found it sufficient to use
a grid resolution of $150 \times 300$ cells. The discussion of the non-linear
behaviour is based on calculations with grid size $400 \times 800$, to
ensure convergence.
In all cases, the grid is surrounded by
two layers of ghost cells used to impose the boundary conditions.  
The boundary conditions are periodic in the $x$-direction, while all quantities
are extrapolated continuously into the ghost cells  at $y=\pm 1$.
The evolution of the instability is not influenced by the boundary
conditions at $y=\pm 1$, as these boundaries 
are sufficiently far away from the central 
region of shear flow. 

\begin{figure}[hb]
 \FIG{\vspace{7cm}
 \includegraphics{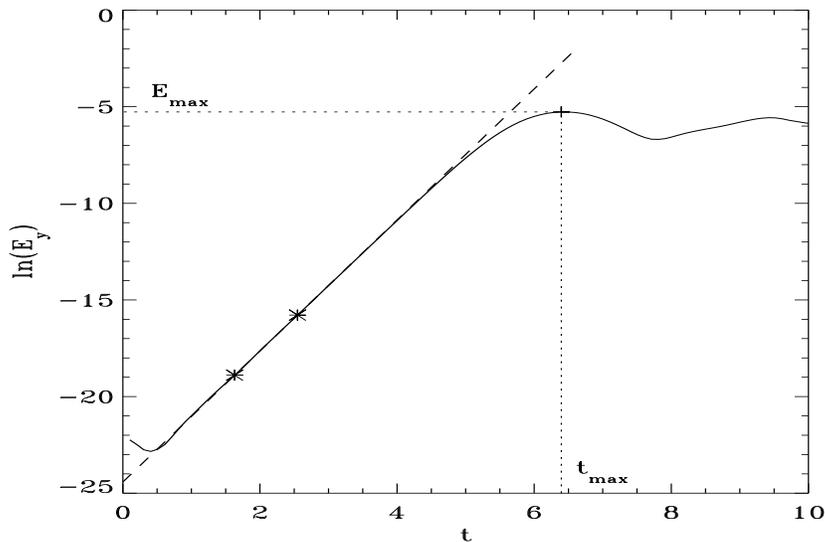}}
 \caption{The time evolution of the (logarithm of the)
total vertical kinetic energy $E_y(t)$. This quantity is 
used to determine the growth rate $\Gamma$ and
the saturation level $E_{max}$, which is reached at time $t_{max}$. 
The growth rate is determined by the linear fit
in this semi-logarithmic plot to the time interval $[0.25 t_{max},0.4 t_{max}]$,
shown by the dashed line.}
\label{f-growth}
\end{figure}

\section{Linear results: growth rates}\label{s-LR}

The initial phase of the evolution is one of exponential growth
in accord with Fig.~\ref{f-khstruc}. A plasma circulation
sets in, forming a periodic pattern of density depletions and enhancements.
We determine the growth rate $\Gamma$ of the KH instability by monitoring the
vertical kinetic energy $E_y= \int\int dx\,dy(\rho v_y^{2})/2$.
The saturation level $E_{max}$ is determined as the first maximum in
$E_y(t)$, which is reached at time $t_{max}$ (see Fig.~\ref{f-growth}).
We then fit $E_y(t)$ 
with an exponential $\exp(2\Gamma t)$ in the time interval $t\in [0.25 t_{max},
0.4 t_{max}]$. As we always initiate the instability using a
small velocity perturbation 
$\delta v_y \simeq {\cal{O}}(10^{-4})$, this time $t_{max}$
is typically within $t_{max} \in [6,11]$. Due to our normalizations, time is 
essentially measured in the transverse sound travel time, $L_y/c_s$. 

\begin{table}
\begin{center}
\begin{tabular}{ccccccccccc}
\vphantom{\LARGE B}
$V_0$ & $a$ & $v_{y0}$ & $\sigma$ & $k_x$ & $M_s$ & $B_0$ & $M_a$ & $\beta$ & $\Gamma a/V_0$ & $2 E_{max}/\rho_0 V_0^2$ \\
\hline
\hline
\vphantom{\LARGE B}
0.645 & 0.05 & $10^{-4}$ & 0.2 & $2\pi$ & 0.5 & 0 & $\infty$ & $\infty$ & 0.134 & 0.038
\\ \hline
\vphantom{\LARGE B}
0.645 & 0.05 & $10^{-4}$ & 0.2 & $2\pi$ & 0.5 & $+0.129$ & 5.0  & 120.2 & 0.131 & 0.025
\\ \hline
\vphantom{\LARGE B}
0.645 & 0.05 & $10^{-4}$ & 0.2 & $2\pi$ & 0.5 & $\pm 0.129$ & 5.0  & 120.2 & 0.185 & 0.033
\\ \hline
\end{tabular}
\caption{Parameters, growth rates, and saturation levels 
for the reference cases.}
\label{t-reference}
\end{center}
\end{table}

\begin{figure}[ht]
\FIG{\vspace{9cm}
 \includegraphics{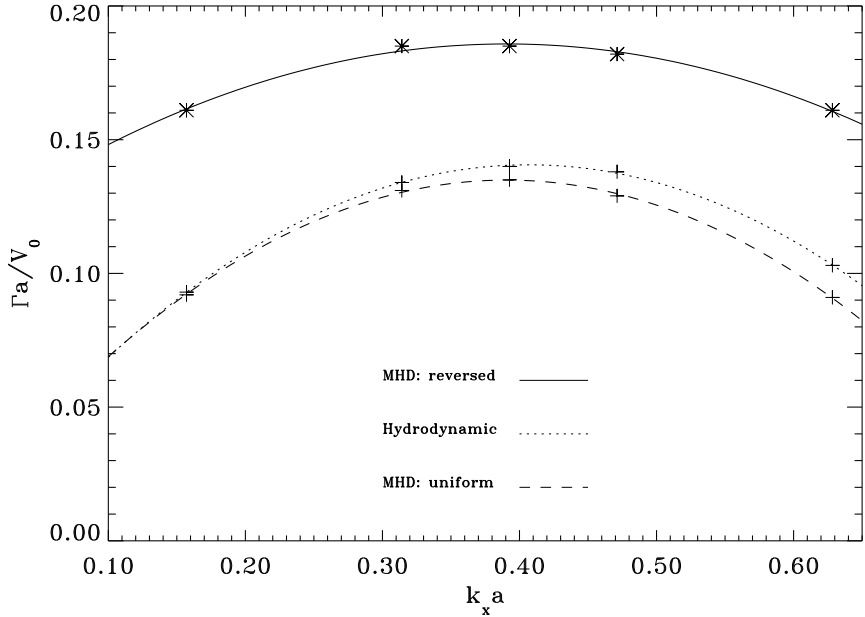}}
 \caption{Growth rate $\Gamma$ versus wavenumber $k_x$. Five calculated
growth rates (`$+$' and `$*$') 
are fitted by a parabola for each case: pure hydrodynamic
(dotted), uniform MHD (dashed), and reversed MHD (solid). Note the
stabilizing effect of a uniform magnetic field, while a reversed field
destabilizes.}
\label{f-wavenumber}
\end{figure}

\begin{figure}[ht]
\FIG{\vspace{9cm}
 \includegraphics{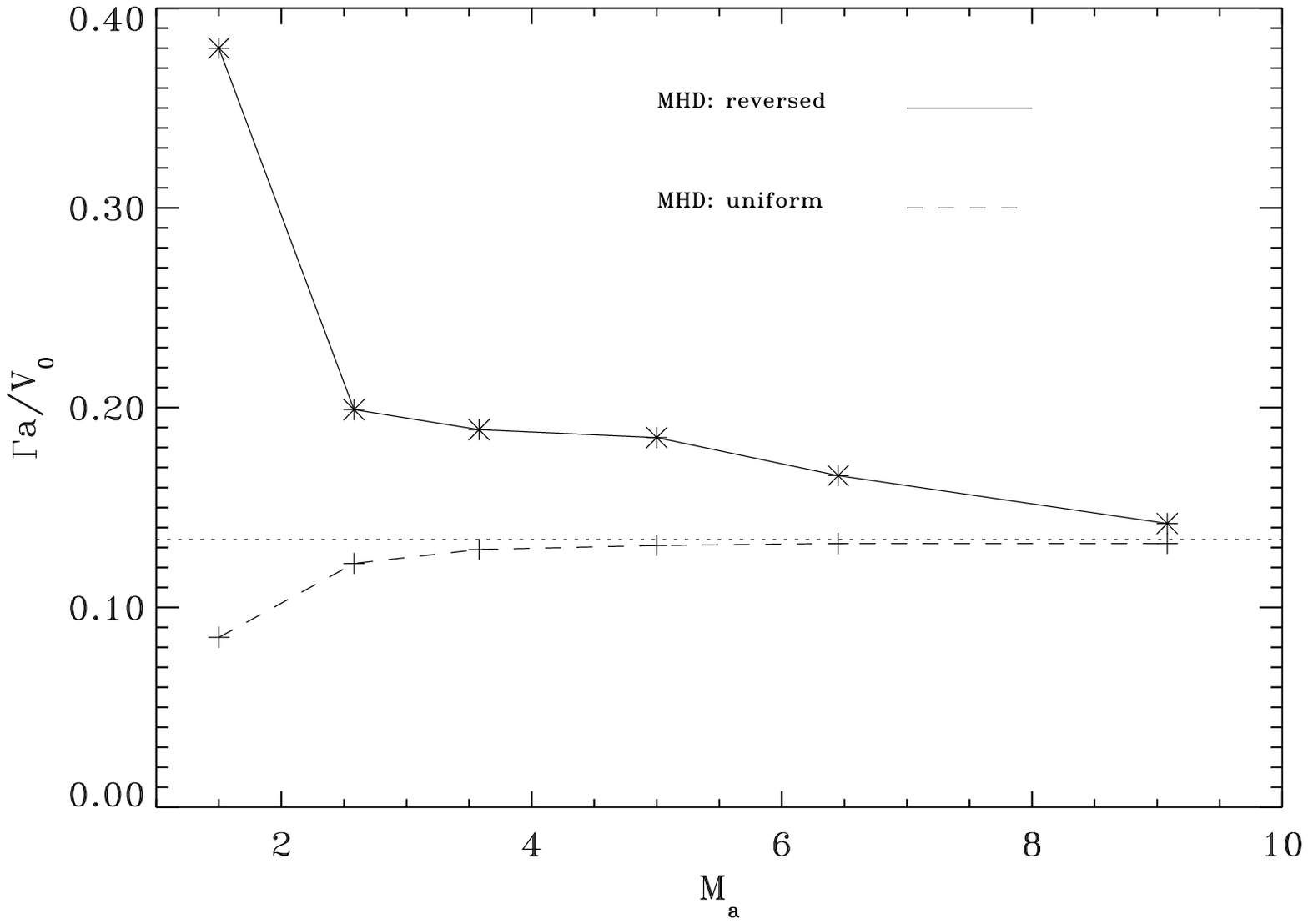}}
 \caption{Growth rate versus Alfv\'{e}n Mach number.
For uniform MHD (`$+$' and dashed line) and reversed
MHD case (`$*$' and solid line). Both reach the hydrodynamic growth
rate (dotted line) at right, in the limit of weak initial magnetic field.}
\label{f-alfvenmach}
\end{figure}

We introduce a reference hydrodynamic case
with wavenumber $k_x=2\pi$ and $V_0=0.645$, such that the sound Mach
number is $M_s=0.5$. With the above procedure to determine its growth rate,
we find $\Gamma a/V_0 = 0.134$. Adding an initially weak (uniform or reversed)
magnetic field $B_0=0.129$ provides us with reference cases for 
MHD situations where $M_a=5.0$ and the plasma beta $\beta=2 p_0/B_0^2 \simeq 
120$.  Under these parameters, the uniform case has a growth rate of 
$\Gamma a /V_0=0.131$, while the reversed case yields $\Gamma a /V_0=0.185$. 
Fig.~\ref{f-growth} corresponds to the uniform reference case.
Consistent with the results of~\cite{miupr} and~\cite{malag}, 
a uniform magnetic field stabilizes
the KH instability. Interestingly, starting from a reversed 
magnetic configuration, we find an accelerated growth. 
We list all parameters specifying the reference 
hydrodynamic, uniform MHD, and reversed MHD case 
and their calculated growth rates and saturation levels 
in table~\ref{t-reference}.
In addition to the listed values, 
we have $L_y=1=L_x$, $\rho_0=1$, $p_0=1$, and $\gamma=5/3$.

\begin{figure}[ht]
\FIG{\vspace{9cm}
 \includegraphics{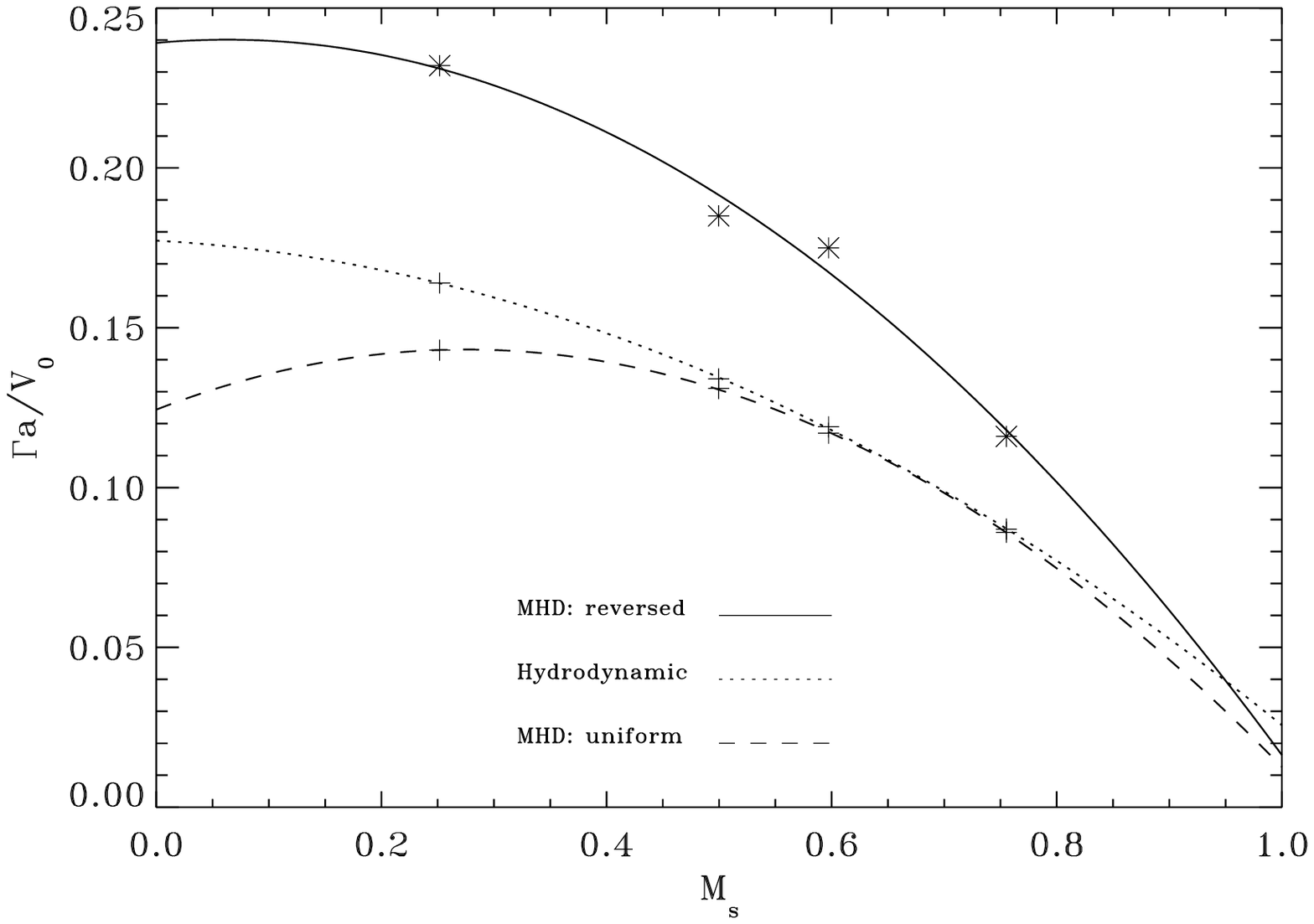}}
 \caption{Growth rate $\Gamma$ versus sound Mach number. A parabolic
fit connects four data values for each case. Beyond $M_s>1$, the 
KH instability is stabilized.}
\label{f-soundmach}
\end{figure}

We determined the dependence of the growth rate on 
the wavenumber $k_x$, and the results are summarized in Fig.~\ref{f-wavenumber}.
We took $k_x=[\pi,2\pi,2.5\pi,3\pi,4\pi]$ with all other parameters as in
the reference cases. The values for the HD case and the uniform MHD case
can be fitted perfectly by a parabola, 
as expected from Fig.~$4$ in~\cite{miupr}. 
The only difference with their Fig.~$4$ is a difference in scaling. 
Two effects are immediately apparent from our Fig.~\ref{f-wavenumber}.
First, the KH instability is stabilized at small and large wavelengths, so that
a maximal growth rate, here at $k_x a=0.4$, is observed.
Reasoning from Fig.~\ref{f-khstruc}, at
small wavenumbers, the distance between the periodic depletions and 
enhancements of the density widens,
leading to a smaller vertical pressure gradient and a more stable situation.
In the other limit of large wavenumber perturbations, the
resulting small circulations of matter cover only a part of the inner 
region $|y|<a$, so the driving force of the instability becomes less effective. 
A stable situation is reached again.    
Second, while a uniform magnetic field reduces the instability at all
wavenumbers, the reversed field acts to destabilize.
This is consistent with the results in the incompressible case given by
\cite{dahl97} and~\cite{dahl98}.

This effect is even more apparent when we fix the wavenumber to $k_x=2\pi$,
and gradually increase the magnetic field, as shown in Fig.~\ref{f-alfvenmach}.
For low field strength (we took $B_0=0.071$, corresponding to an Alfv\'en
Mach number $M_a=9$ at very high $\beta=397$), the hydrodynamical limit of the 
growth rate at $k_x a=0.314$ (reference case)
is reached at the right edge of the figure. 
Increasing the magnetic field strength up to $B_0=0.43$ where $M_a=1.5$,
we clearly demonstrate the stabilizing effect of a uniform magnetic
field and the destabilizing effect of a reversed initial field.
In fact, for $B_0=0.645$ or $M_a=1$, the uniform case is completely
stabilized, while the reversed case is not.
The stabilizing effect of a uniform magnetic field parallel to the
shear flow is due to magnetic tension: in ideal MHD, the imposed
perturbation $\delta v_y$ which is perpendicular to the initial field
lines entrails the field lines and builds up the restoring
magnetic tension. At the same time, field lines are pushed closer
together in those regions at the interface
where the flow converges and thereby enhances
the density. It is this effect which, in turn, destabilizes the reversed
field case. Indeed, when forcing anti-parallel field lines towards
one another, (numerical) diffusion will allow for reconnection in those
places, so that the magnetic tension in the reconnected field
lines adds to the pressure gradient force indicated 
in Fig.~\ref{f-khstruc}. Therefore, a faster growth ensues.
At the grid resolution $150 \times 300$ used to determine
these growth rates, the numerical diffusion inherent in the
scheme used is similar to a resistivity of about $\eta\simeq{\cal{O}}(10^{-5})$.
This value is found by comparing the ideal MHD
calculations with resistive MHD runs with $\eta=10^{-4},\, 3\times 10^{-5}, \,
10^{-5}$. In the high resolution non-linear calculations 
discussed below, we will therefore
solve the resistive MHD equations with a non-zero resistivity of $\eta=10^{-5}$. 
Since the resistivity is physically significant in the reversed
field cases, we must address
the limitations associated with the initial discontinuous
magnetic field profile. We conducted experiments with an 
initial $\tanh(y/b)$-profile
for $B_x$ with $b=0.01$ in combination with a grid accumulation
about $y=0$. For the reference magnetic field strength of
$B_0=0.129$, the growth rate found when starting from this continuous
initial current sheet then reduces to $\Gamma a/V_0=0.127$, even
lower than the reference uniform MHD case! This reduction is stronger
for larger field strengths. This stabilization is related to
the fact that for this continuous case, the initial thermal pressure
profile necessarily peaks at the shear layer in proportion to $\sim 1+B_0^2/2$.
This has a significant influence on the development of the KH instability
as can be expected from Fig.~\ref{f-khstruc}.
However, the linear behaviour in the limit of $b\rightarrow 0$ is 
well represented by Fig.~\ref{f-alfvenmach}, at least for field
strengths comparable or weaker than the reference case. Indeed, the initial
discontinuous profile rapidly smears out over a few grid cells by the numerical
method, so it effectively reduces to the continuous case for small $b$.
Moreover, the non-linear behaviour of the reference reversed case
for $b=0$ versus $b=0.01$ is almost unaltered, as discussed below.

To conclude the linear results, we show the dependence of the
growth rate $\Gamma$ on the sound Mach number in Fig.~\ref{f-soundmach}.
Starting from the reference cases, we 
considered values for the background shear flow given by
$V_0=[0.325,0.645,0.771,0.975]$. Note that this also results in
variations of the Alfv\'en Mach number for the MHD cases from
$M_a=2.5$ up to $M_a=7.6$. Again, we can see 
the extra elastic properties of the uniform background magnetic field 
causing the overall decrease in growth rate,
and the destabilizing effect of a reversed field.
In all cases, the growth rate decreases for increasing sound Mach number.
As we increase the background shear flow, 
the amount of kinetic energy in the perturbation becomes smaller, 
relative to the basic flow energy. A more stable situation results as less
energy is available to do work for compressing the fluid.
The dotted curve for the hydrodynamic case is analoguous with Fig.~$1$ 
in~\cite{blumen}. It corresponds to the dotted line
in that figure, connecting the isolines of the growth rate.     

\section{Non-linear results: saturation and further evolution}\label{s-NLR}

We explained how we determined growth rates by fitting the phase
of exponential growth in the vertical kinetic energy $E_y(t)$. This
quantity eventually reaches a maximum $E_{max}$ which we now use as a
measure of the saturation of the KH instability (see Fig.~\ref{f-growth}). 
We normalize $E_{max}$ with the kinetic energy corresponding to the
initial shear flow $\rho_0 V^2_0 /2$. Note that additional,
possibly higher, maxima in $E_y(t)$ may occur later on.
In practice, a low value for the first maximum (say $<0.01$) 
clearly indicates that
the KH instability is strongly suppressed by non-linear effects.

\begin{figure}[ht]
\FIG{\vspace{9cm}
 \includegraphics{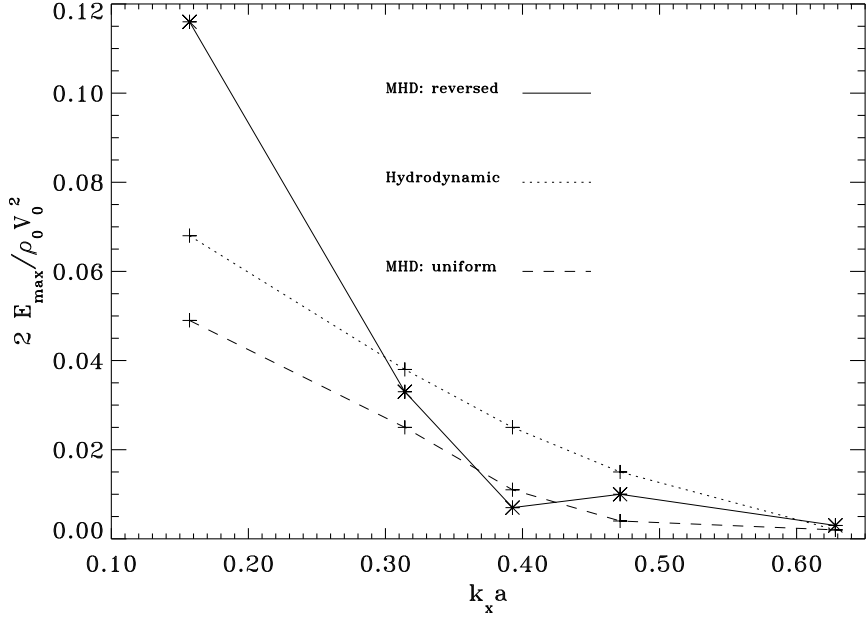}}
 \caption{Saturation level $E_{max}$ versus wavenumber $k_x$. Compare with
the dependency of the linear growth rate shown in Fig.~\ref{f-wavenumber}.}
\label{f-Swavenumber}
\end{figure}

\begin{figure}[ht]
\FIG{\vspace{9cm}
 \includegraphics{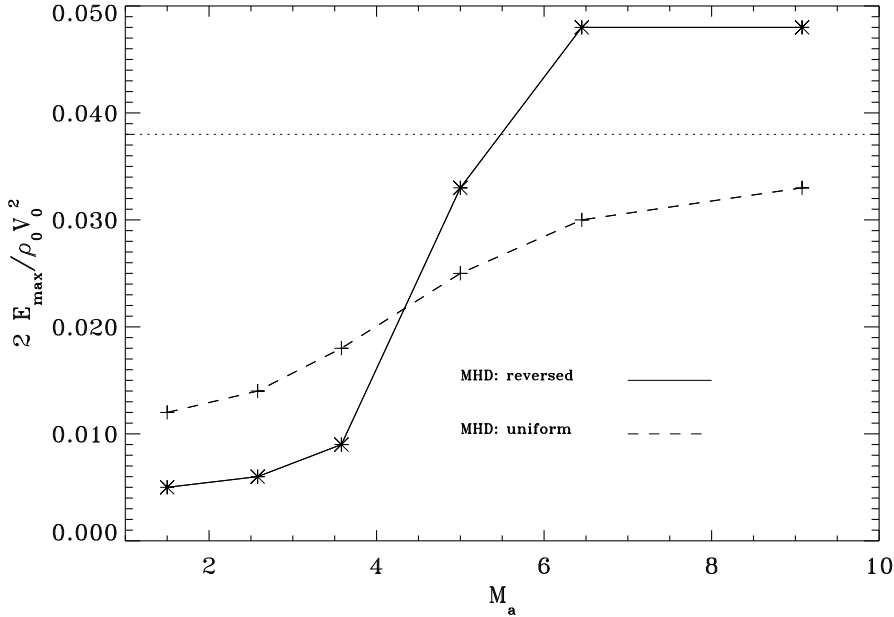}}
 \caption{Saturation level versus Alfv\'{e}n Mach number. A similar
pure hydrodynamic case saturates at $2E_{max}/\rho_0 V^2_0\simeq 0.038$
(horizontal dotted line),
so a uniform magnetic field is non-linearly stabilizing.}
\label{f-Salfvenmach}
\end{figure}

\begin{figure}[ht]
\FIG{\vspace{9cm}
 \includegraphics{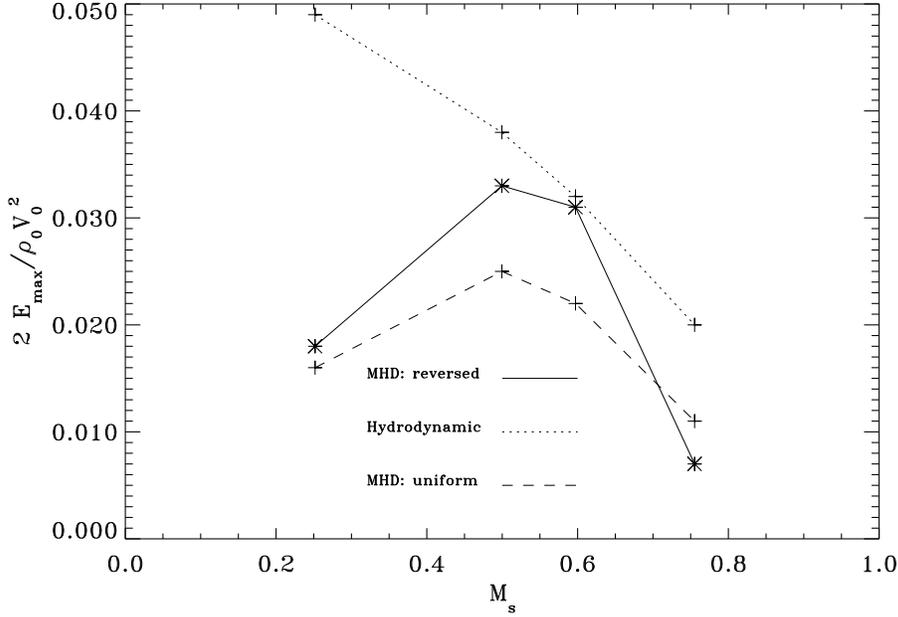}}
 \caption{Saturation level versus sound Mach number. For $k_x=2\pi$ and
$B_0=0.129$, both MHD cases saturate below the pure hydrodynamic
case for the range of shear flows $0.325 \leq V_0 \leq 0.975$.}
\label{f-Ssoundmach}
\end{figure}

\begin{figure}[ht]
\FIG{\vspace{17cm}
 \includegraphics{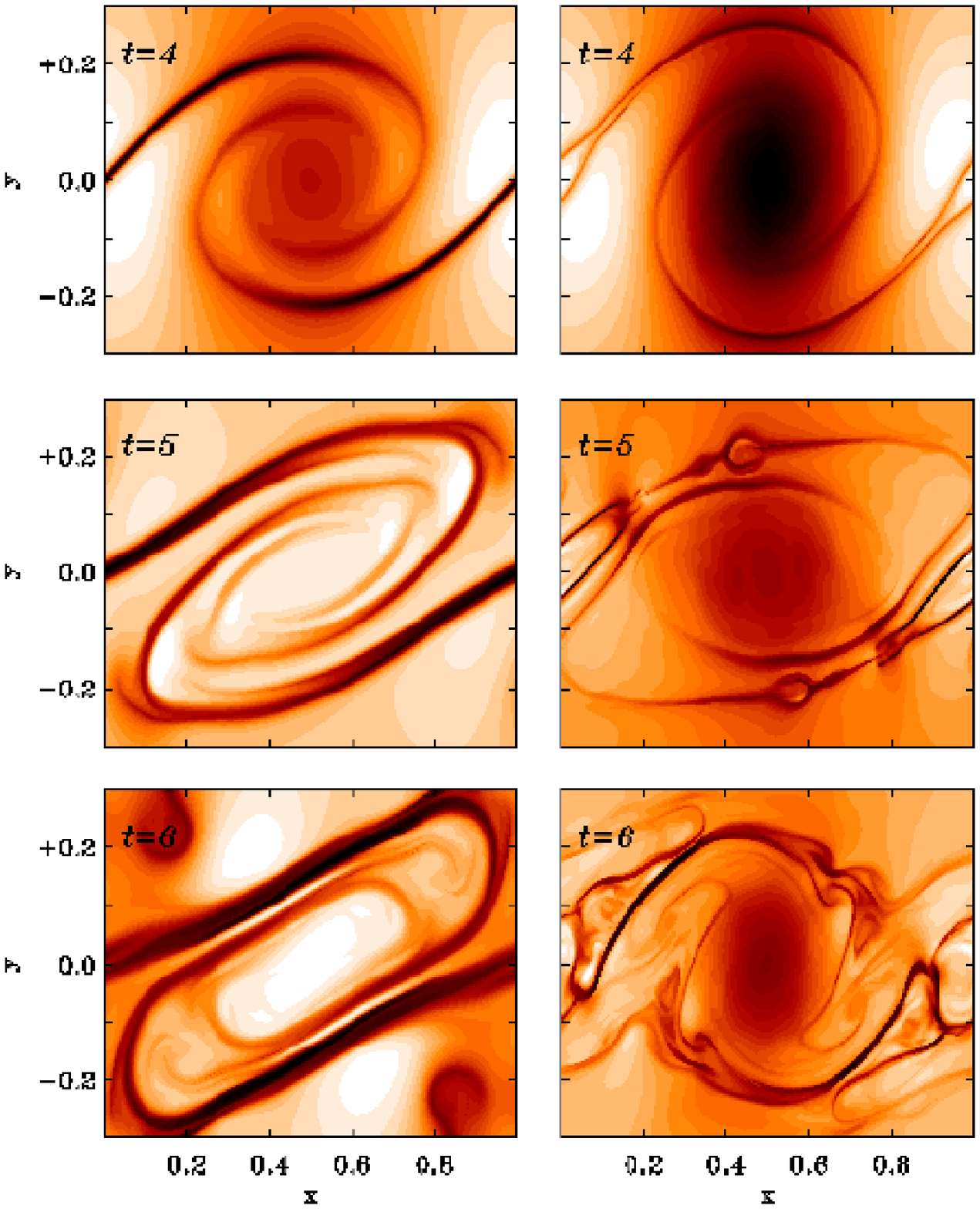}}
 \caption{Evolution of the density at times $t=4$ (top), $t=5$ (middle), 
and $t=6$ (bottom), for the 
case of a uniform (left frames) and of a reversed field (right frames).
We took $k_x=2\pi$, $V_0=0.645$, and $B_0=0.129$. The non-linear development
differs markedly.}
\label{f-density}
\end{figure}

Figures~\ref{f-Swavenumber},~\ref{f-Salfvenmach}, and~\ref{f-Ssoundmach}
show the dependence of the saturation level $E_{max}$ on the wavenumber,
Alfv\'{e}nic Mach number, and sound Mach number, respectively. 
Note how the saturation level for the hydrodynamic and the uniform MHD case
monotonically increases with decreasing wavenumber in
Fig.~\ref{f-Swavenumber}, while we demonstrated in
Fig.~\ref{f-wavenumber} that the growth rate has a pronounced
maximum in the range of wavenumbers considered.
As a consequence, the linearly fastest growing mode does
not necessarily correspond to the dominant mode in the non-linear evolution.
However, the stabilizing effect
of a uniform magnetic field persists in the non-linear regime
(see~\cite{frank} and~\cite{malag}), as we
find saturation levels which are always lower than in a pure HD case.
Part of the available energy must be used to compress and stretch the
field lines, leading to a lower $E_{max}$.
For the reversed field case, no clear trend with wavenumber is
apparent, and the KH instability can saturate at intermediate,
lower or higher levels when compared to pure HD or uniform MHD cases.
We point out that the reversed case is most susceptible to the inherent
numerical diffusion. By calculating reversed cases on
larger grids, and by investigating the limiting $\eta \rightarrow 0$ behaviour,
we are confident that the observed variation is real. Nevertheless,
the absolute values for the saturation level for the reversed field cases
are less certain than the ones shown for hydrodynamic and uniform MHD cases.
The saturation levels for the reversed case are fairly insensitive to
the specific initial $B_x$ profile: the reference case with an initial
discontinuity at $y=0$ saturated at $2 E_{max}/\rho_0 V_0^2 \simeq 0.033$,
while starting from a $\tanh(y/b)$ with $b=0.01$, the saturation was reached
at $\simeq 0.039$. The qualitative non-linear behaviour is identical.

The dependence of the saturation
level on Alfv\'enic Mach number
at fixed $k_x=2\pi$ is shown in Fig.~\ref{f-Salfvenmach}. The hydrodynamic
saturation level is situated at $2 E_{max}/\rho_0 V_0^2 \simeq 0.038$ 
(horizontal dotted line),
and the uniform MHD case approaches this value from below when
the initial field strength is decreased. However, the reversed case 
saturates sooner than the uniform MHD case at initial field strengths
$B_0 \geq 0.18$, while for $B_0 \leq 0.1$, it saturates above the HD limit.
Varying the shear flow $V_0$ at fixed wavenumber $k_x=2\pi$ 
as in Fig.~\ref{f-soundmach} leads to a similarly complicated dependence of the
saturation level on the sound Mach number $M_s$. Fig.~\ref{f-Ssoundmach}
demonstrates how at this wavenumber and for fixed initial field
strength $B_0=0.129$, both the uniform and the reversed MHD cases
saturate at a lower level than a pure hydrodynamic situation.

To gain insight in the non-linear behaviour of the KH instability in
both uniform and reversed MHD cases, we confront in Fig.~\ref{f-density}
time series of the density pattern for both. We took
the initial parameters as in the reference cases ($k_x=2\pi$, $V_0=0.645$
and $B_0=0.129$), but as our interest is now in the non-linear regime
only, we took $v_{y0}=0.01$ so that the non-linear stage can be reached
at a lower computational cost. The uniform case at left was calculated in
ideal MHD on a $300 \times 600$ grid, while the reversed case at right
is done in resistive MHD with $\eta=10^{-5}$ and the use of $400\times 800$
grid points. We show the part of the grid between $-0.3 \leq y \leq 0.3$.
The contour levels differ from frame to frame, as we renormalized
individual colour ranges by the instantaneous extremal values of the density
to bring out all detail. The actual range of density values over all
frames plotted is $0.58 \leq \rho(x,y,t) \leq 1.15$.

The frames for the uniform MHD case are in good
agreement with Ma\-la\-go\-li et al. (1996).
%~\cite{malag}.
These authors concluded that the evolution of the KH instability in the
presence of an initially uniform field consists of three phases.
First, the instability grows exponentially, very much like 
Fig.~\ref{f-khstruc}. 
Second, the density pattern becomes completely controlled 
by the magnetic field which is strengthened in the process and becomes
dynamically important. Indeed, field lines are pushed away from the center 
of the layer and stretched by the vortical flow. The first snapshot
shown in Fig.~\ref{f-density} corresponds to the time of saturation where
the dark lanes in the regions of enhanced density (bright) outline these
stretched and compressed field lines. From the moment where the gradient
scales across these field lines diminish to trigger
(numerical) dissipation in this ideal MHD run, reconnection events occur.
The second and third snapshot shows the density pattern in the first
stages of this regime. 
At that point, the magnetic energy has already reached its maximum value, 
and starts to decay 
oscillatory. The third and final phase is one where small scale turbulence 
develops, until a new statistically-steady flow sets in. 
We have continued our calculation, and in agreement
with~\cite{malag}, we end up with an enlarged,
mixed layer directed along the initial shear flow, aligned with
the magnetic field (not shown).

The frames at right for the reversed case are dramatically different.
We already mentioned that even in the linear regime, flux cancellation
at $y=0$ sets in almost instantly.
In fact, the reversed MHD case starts to resemble a pure HD case more
closely: the top right panel ($t=4$)
shows how a central density depletion (dark) 
forms since the dissipated field there no longer controls the dynamics.
The effects of the magnetic field are most pronounced
at some distance away from the initial interface at $y=0$, and at those
regions where field lines are pushed together.
Already at the first frame shown at $t=4$, an island structure is
evident within the region of enhanced density (bright). 
In contrast to the uniform
case, now anti-parallel field lines are pushed together there. 
Such a situation is tearing unstable in resistive MHD, and leads
to the formation of a magnetic island, which influences
the density dynamically to form the pattern shown.
This process continues along the current sheet, as seen in the
second frame where more, smaller islands are evident. Reconnection
and tearing thus occurs all along the current sheet as it gets
amplified by the KH vortical flow. 

\begin{figure}[ht]
\FIG{\vspace{10cm}
 \includegraphics{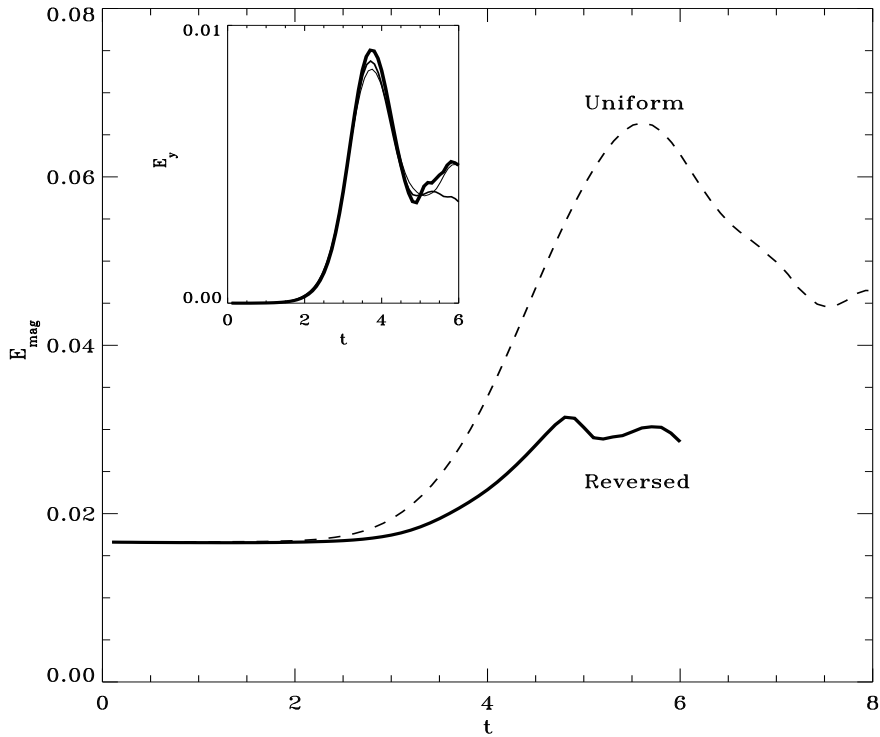}}
 \caption{Evolution of the magnetic energy $E_{mag}(t)$ for the 
case of a uniform (dashed) and of a reversed field (solid).
The uniform case is calculated in ideal MHD on a $300\times 600$ grid.
The reversed case is a resistive MHD run with $400 \times 800$ grid points.
The inset shows the evolution of the vertical kinetic energy
$E_{y}(t)$ for resistive $\eta=10^{-5}$ MHD runs on $150 \times 300$,
$300 \times 600$ and $400\times 800$ grids.}
\label{f-magnetic}
\end{figure}

\begin{figure}[ht]
\FIG{\vspace{9cm}
 \includegraphics{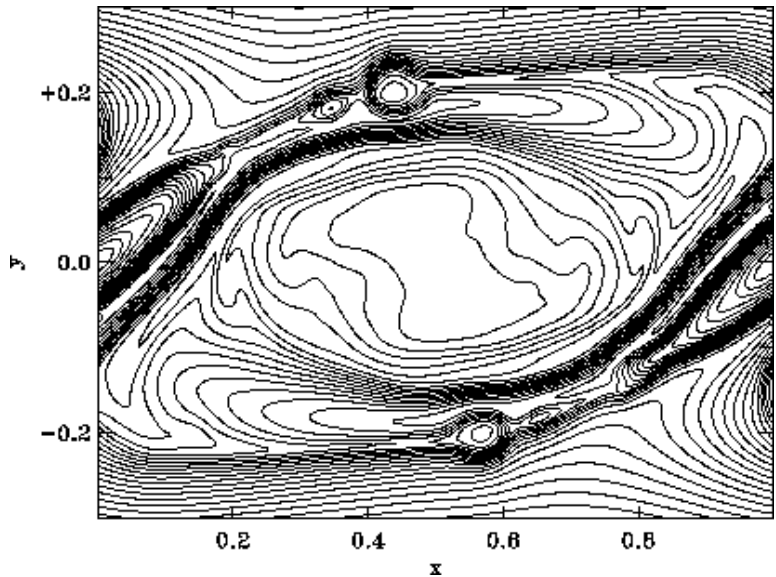}}
 \caption{Island formation due to tearing instability for a reversed
MHD case. We show
the magnetic field structure at $t=5$, corresponding to the middle
right frame of Fig.~\ref{f-density}.}
\label{f-bfield}
\end{figure}

Naturally, the initially
reversed field is thereby dissipated faster than in the uniform MHD case.
A crude monitor for this is given by the time evolution
of the total magnetic
energy $E_{mag}(t)=\int\int\,\, dx\,dy(B_x^{2}+B_y^{2})/2$ for both cases.
This is shown in Fig.~\ref{f-magnetic} where the uniform
case (dashed line) is compared to the reversed case (solid line). 
Note how the maximal magnetic energy reached in the reversed case is
much lower than for a uniform background field of equal strength.
The inset shows the result of a convergence study
in resistive MHD for the reversed case only: fixing
$\eta=10^{-5}$, we compare calculations on grids of
size $150\times 300$, $300\times 600$ and $400\times 800$. 
The time evolution of the vertical
kinetic energy $E_y(t)$, used to determine the saturation levels earlier,
is clearly captured on all grids up to $t=5$.
The magnetic field pattern at $t=5$ for the reversed
case is shown in Fig.~\ref{f-bfield}. The islands coincide
with the density perturbations visible in Fig.~\ref{f-density} (middle
right panel). Beyond this time, small-scale turbulence sets in. This is evident
from the last panel (time $t=6$) shown in Fig.~\ref{f-density}.

\begin{figure}[ht]
\FIG{\vspace{9cm}
 \includegraphics{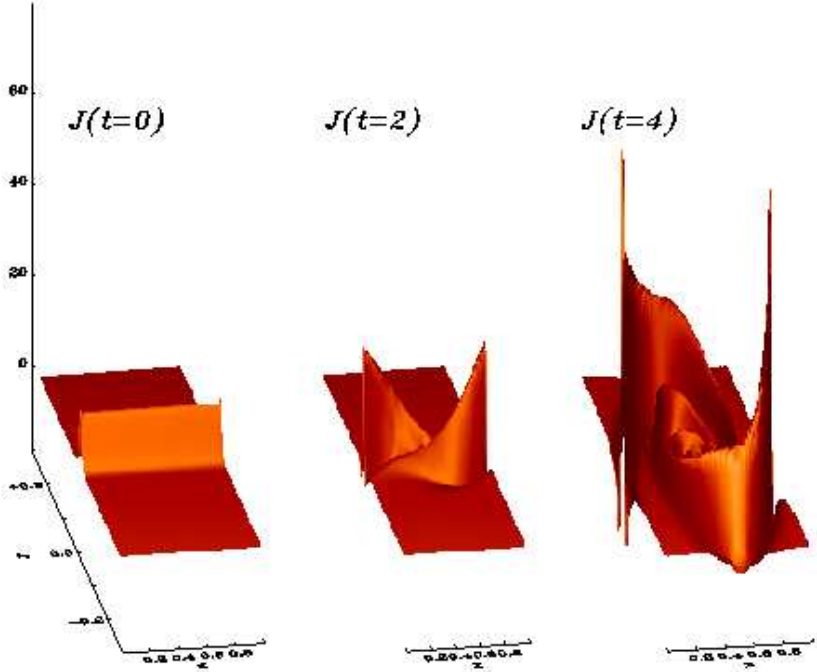}}
 \caption{The amplification of an initially continuous current sheet in
a KH unstable shear flow. Apart from the finite extent of
the current sheet at $t=0$, parameters are identical to our
reference reversed case shown at left in Fig.~\ref{f-density}.}
\label{f-current}
\end{figure}

To demonstrate that the island formation correctly represents the non-linear
development of combined current-vortex sheets (initial $\tanh$ profiles for
{\em both} $v_x$ and $B_x$), we show in Fig.~\ref{f-current}
the amplification of an initial
continuous current sheet with $b=0.01$ by the developing vortex flow. 
All other parameters are as in the reference reversed case shown in 
Fig.~\ref{f-density}, but now we took a $150\times 300$ grid with
grid accumulation about $y=0$. The last frame shown corresponds
with the top right panel of Fig.~\ref{f-density}. The island structure
is cospatial with the pronounced current maximum.

\section{Discussion and conclusions}\label{s-C}

We studied the KH instability both in
hydrodynamics and in magnetohydrodynamics with the
Versatile Advection Code (VAC). 
We investigated the linear and the non-linear
saturation regime by varying wavenumbers,
sound Mach numbers, and, for the uniform and reversed field MHD case, 
the Alfv\'enic Mach number. While non-linear results for
the hydrodynamic and uniform MHD cases confirm and extend results by
previous authors, we demonstrated that
novel physical effects appear when a reversed magnetic field
is imposed.

The linear results clearly show the stabilizing effect of the uniform 
magnetic field, leading to smaller growth rates than the
hydrodynamic case. This stabilizing effect is also present in the non-linear
regime, since the strengthened, initially uniform field eventually halts 
the exponential growth sooner. This phase where the magnetic field controls
the density pattern, eventually decays into a turbulent regime, as previously
pointed out by~\cite{malag}. We focused on the
saturation behaviour of the KH instability, and discussed its dependence
on wavenumber and Mach numbers. 

When the shear flow coincides with a field reversal, reconnection takes
place right away, so our results for the reversed case 
are applicable in resistive MHD. We started from a situation where the
field reversal is modeled as a discontinuity, seen as the limiting case of
cospatial current-vortex sheets 
where the region of field reversal is completely contained
within the region of velocity shear (see also~\cite{dahl97,dahl98}). 
In the linear regime,  
we find an overall larger growth rate than in hydrodynamic and uniform
magnetohydrodynamic cases. Hence, a reversed field destabilizes the
KH instability, and when the strength of
the reversed magnetic field decreases, we approach the hydrodynamic
growth rate from above. 
Comparisons with cases where the field reversal is a continuous $\tanh$-profile
revealed that this linear destabilization is sensitive to the initial condition.
Only for moderate magnetic fields (like our reference reversed case),
the destabilizing effect persists when the region of magnetic shear is
significantly narrower than the region of velocity shear. 
The role played by the initial pressure profile is crucial for stronger
and wider current sheets.
Such strongly magnetized, wide current sheets have been the subject of
analytic studies of current-vortex sheets in 
incompressible MHD (\cite{dahl97}). There, the ratio of the magnetic shear width
to the velocity shear width ($b/a$ in our terminology) was typically taken
around unity and the Alfv\'en Mach number was varied from 0.6 to 1.2.
We restricted our study to cases where $b \ll a$ and $M_a > 1$, since it is
known that the uniform MHD case is fully stabilized when $M_a=1$. Our study
indicates (cfr. the evolution of the density in Fig.~\ref{f-density}) that
compressibility plays an important role in current-vortex sheet
dynamics. We focused on KH unstable configurations where the magnetic effects
are secondary. To make the connection with the results of~\cite{dahl97} more
explicit, we also calculated one case with $b=0.2 a$ and $M_a=1$, setting
$\eta=10^{-3}$, very much like Case I of their study. In
our compressible simulation, the density varied locally up to
40 \% during the development and the saturation of the then dominant
tearing instability. A quantitative comparison with the incompressible
simulations is therefore fairly complicated. 

In our reversed case study, reconnection sets in almost instantly at
the shear flow interface and the reconnected field lines act to enforce the 
KH instability.
Flux cancellation at $y=0$ clearly alters the field structure earlier 
than in a similar field-strength uniform case.
While the uniform magnetic field effectively prevented 
the build up of a central density perturbation, by dynamically influencing 
the flow pattern throughout the inner region, 
the field in the reversed case changes the flow and density most markedly
at some distance away from $y=0$.
Reconnection is also important in those regions where anti-parallel field
lines are pushed together, thereby amplifying the initial current sheet. We showed
how the KH flow may thus in turn drive this sheet unstable. Resistive tearing
modes form magnetic islands 
and influence the density pattern by isolating blobs of enhanced
density. Decay to a turbulent regime sets in sooner than for a uniform
MHD case of equal field strength.

The temporal evolution of the total 
magnetic energy shows a much lower maximal energy level in the case of the
reversed field compared to the uniform field case. 
The saturation as measured by the first maximum in the vertical kinetic energy 
indicates
that the reversed field can saturate at a lower, a higher, or at an
intermediate level between a
pure hydrodynamic and a uniform MHD case.
The temporal evolution of the 
density in the reversed field case initially resembles 
the hydrodynamic case more closely. 

The plane-parallel configuration we investigated here is relatively simple,
however, we expect that the basic features of the KH 
instability are well-described in our model.
Meaningful extensions would model the evolution in three dimensions. A
first step is found in~\cite{miura}, where a magnetic field in the plane 
perpendicular to the velocity gradient is considered.
There, it was found that the vortex train formed by the KH instability
is further susceptible to vortex pairing following the non-linear saturation. 
For fluid applications, the surface tension can not be ignored. 
For certain astrophysical situations, an accurate treatment
of the KH instability must include the influence of viscosity, gravity,
and relativistic flows.

\begin{acknowledgments}
\textit{Acknowledgements.}
The Versatile Advection Code was developed as part of the project on
`Parallel Computational Magneto-Fluid Dynamics', funded by the
Dutch Science Foundation (NWO)
Priority Program on Massively Parallel Computing, and
coordinated by Prof. Dr. J.P. Goedbloed. Computer time on the
Cray C90 was sponsored by the Dutch Stichting
Nationale Computerfaciliteiten (NCF).
G.T. thanks the Astronomical Institute at Utrecht for its hospitality
during his work there.
He is currently supported by the postdoctoral fellowship D25519 of
the Hungarian Science Fundation (OTKA).
\end{acknowledgments}

\end{document}